\documentclass[10pt,a4paper,twoside]{article}
\usepackage{epsfig}
\usepackage{baltlat6}
\usepackage{wrapfig}
\begin{document}
\ \
\vspace{-0.5mm}

\setcounter{page}{1}
\vspace{-2mm}

\titlehead{Baltic Astronomy, vol.\ts 16, xxx--xxx, 2007.}

\titleb{ESTIMATION OF TIME DELAYS FROM TWO BLENDED LIGHT CURVES OF
GRAVITATIONAL LENSES}

\begin{authorl}
\authorb{A.~Hirv}{},
\authorb{T.~Eenm\"ae}{},
\authorb{L.~J.~Liivam\"agi}{} and
\authorb{J.~Pelt}{}
\end{authorl}

\moveright-3.2mm
\vbox{
\begin{addressl}
\addressb{}{Tartu Observatory, T\~{o}ravere, 61602, Estonia}
\end{addressl}
}

\submitb{Received 2007 March 28; accepted 2007 June 14}

\begin{summary} Long time photometric monitoring programs of
gravitational lens systems are often carried on using modest
equipment.  The resolution of such observations is limited and some of
the images may remain unresolved.  It may be still possible to find a
full set of time delays from such a blended data.  We discuss here a
particular but interesting case when we have two light curves that both
are blends.  A suitable computational algorithm is developed and tested
to work with computer-generated model light curves.  Our method combines
both blended sequences using the hypothetical time delays between the
initial components of the light curves as free input parameters.  The
combined curves are then compared using statistical distance estimation.
It occurs that using an assumption of equal magnification ratios between
the components of the blends, we can indeed recover the whole set of
time delays.  \end{summary}


\resthead{Time delays from two blended light curves of gravitational
lenses}{A.~Hirv, T.~Eenm\"ae, L.~J.~Liivam\"agi, J.~Pelt}

\sectionb{1}{INTRODUCTION}

To find the time delays caused by differences in light paths of a
gravitational lens system, we need at least some recognizable features
in the observed light curves.  As the longest delays in some systems can
be hundreds of days, we need to have sufficiently long measurement sets.
Long-time monitorings of such systems are feasible with telescopes of
modest size and resolution.  The best example of this kind of photometry
is the long time series obtained by Schild et al.  (1997).  The
constrained resolution can be also a problem for some large scale
photometry programs.  This motivates us to investigate possibility to
recover time delays from the data which are not fully resolved.  The
general scheme of the relevant algorithms was developed in Hirv et al.
(2007).

In the following we will focus on the case of four original images whose
unresolved observations result in two blended light curves.  To recover
the full set of time delays, we will use the principles of computing the
dispersion spectra introduced in Pelt et al.  (1994) and Pelt et al.
(1996).  Hirv et al.  (2007) developed a similar method and applied it
to a three-image system where two original light curves were blended
together but the third curve was fully resolved.

The method of dispersion spectra was singled out as a base for our
algorithms because of its conceptual simplicity.  There are many other
methods available.  For the latest see Burud et al.  (2001), Koptelova
et al.  (2006), Vakulik et al.  (2006), Cuevas-Tello et al.  (2006) and
references therein.  Some of these new methods can be generalized to
handle blended data.  Some critical remarks about the method of
dispersion spectra can be found in Gil-Merino et al.  (2002) and
response to the critique in Pelt et al.  (2002).

There are two simplifying implicit assumptions in our treatment below.
First, we suppose that the data sets contain a substantial number of
observations with sufficient time coverage, and secondly, we totally
ignore possibility of distortions due to the effect of microlensing.
Consequently, when applying the new method to real observational data,
certain care must be exercised.

Our paper is organized as follows.  First we introduce the basic
ingredients of the new algorithm.  Then we present test data generation
methods.  In the next part we describe the results of the numerical
tests and discuss various implementation details which can be of use for
the prospective users of the new method.  The relevant software modules
may be obtained from the authors.

\sectionb{2}{THE METHOD}

\subsectionb{2.1}{The continuous case}

Let us have a quasar image split into four components by an
intervening gravitational lens.  Formally we have four functions of the
quasar source variability $q(t)$:  $f_{k}(t)=a_{k}q(t-t_{k})$,
$k=1,...,4$, where $a_{k}$ are the magnification coefficients and
$t_{k}$ are the {\it flight times} due to different flight paths.  Our
observational equipment can supposedly record only two images as the
close pairs of $f_{1}$, $f_{2}$ and of $f_{3}$, $f_{4}$ are blended
together due to insufficient resolution.  Thus the corresponding signals
$g_{1}(t)$ and $g_{2}(t)$, that we are going to observe, are the
following functions of the source variability $q(t)$:
\begin{equation}
g_{1}(t)=a_{1}q(t-t_{1})+a_{2}q(t-t_{2}),
\label{g1}
\end{equation}
\begin{equation}
g_{2}(t)=a_{3}q(t-t_{3})+a_{4}q(t-t_{4}).
\label{g2}
\end{equation}

As the spatial separation of $f_{1}$ and $f_{2}$ is small, we may
assume, that $a_{1}\approx a_{2}$ and similarly $a_{3}\approx a_{4}$ for
$f_{3}$ and $f_{4}$.  The amplification ratio between $g_{1}(t)$ and
$g_{2}(t)$ is then $a\approx a_{1}/a_{3}$.  Let the time delay between
$f_{1}(t)$ and $f_{2}(t)$ be $\Delta a=t_2-t_1$, and the time delay
between the components of the second image $\Delta b=t_4-t_3$.  These
delays are typically rather short due to nearby flight paths for the
component images.  As the paths of $f_{1}(t)$ and $f_{3}(t)$ differ
significantly (larger spatial separation), the corresponding delay
$\Delta c=t_3-t_1$ is the longest one.  Now we can rewrite the
Eqs.~(\ref{g1}) and~(\ref{g2}) in terms of the first subimage $f_{1}(t)$
and relative time delays:
\begin{equation}
g_{1}(t)=f_{1}(t)+f_{1}(t-\Delta a),
\label{gg1}
\end{equation}
\begin{equation}
g_{2}(t)=f_{1}(t-\Delta c)+f_{1}(t-\Delta c-\Delta b).
\label{gg2}
\end{equation}

To keep things easier to follow we did not multiply the Eq.~(\ref{gg2})
by the amplification ratio $a$.  The fact, that $g_{1}$ and $g_{2}$ may
have different baselines and amplitudes is taken into account in our
computational (matching) algorithm.  As a schematic example of the
initial variability, the $f_{1}(t)$ is shown as a single-peaked function
in Figure~1.  Shifting it by delays $\Delta a$, $\Delta b$ and $\Delta
c$ and adding the results as in the Eqs.~(\ref{gg1}) and~(\ref{gg2}) we
get the double peaked blends $g_{1}(t)$ and $g_{2}(t)$ of the source
variability.

To recover all the three independent time delays $\Delta c$, $\Delta a$
and $\Delta b$ hidden in the light curves $g_{1}(t)$ and $g_{2}(t)$, we
will combine the data using three trial delays $\delta c$, $\delta a$
and $\delta b$ into artificial blends $A(t)$ and $B(t)$:
\begin{equation}
A(t)=g_{1}(t-\delta c)+g_{1}(t-\delta c-\delta b),
\label{at}
\end{equation}
\begin{equation}
B(t)=g_{2}(t)+g_{2}(t-\delta a).
\label{bt}
\end{equation}

If it happens, that $\delta c=\Delta c$, $\delta a=\Delta a$ and $\delta
b=\Delta b$, the difference of $A(t)$ and $B(t)$ vanishes to zero and
this is the situation we are going to search for. The composition of
the
artificial blends $A(t)$ and $B(t)$, when the trial delays match the initial
delays, is also shown in Figure~1.  For clarity we plotted the
components of the artificial blends before and after adding.  Blends and
components that have the same origin are plotted using the same line
type.  As we can see, artificial blends have the same profile, when
trial delays correspond to the initial ones, and the difference between
$A(t)$ and $B(t)$ vanishes.  This is the idea of our method in terms of
the continuous and noise-free light curves.  Next we will see, how this
kind of construction can be used for real noisy and sampled data.  (For
detailed discussion on adding and subtracting of sampled noisy
time-series see Hirv et al. 2007).

\begin{figure}[!t]
\vbox{
\centerline{\psfig{figure=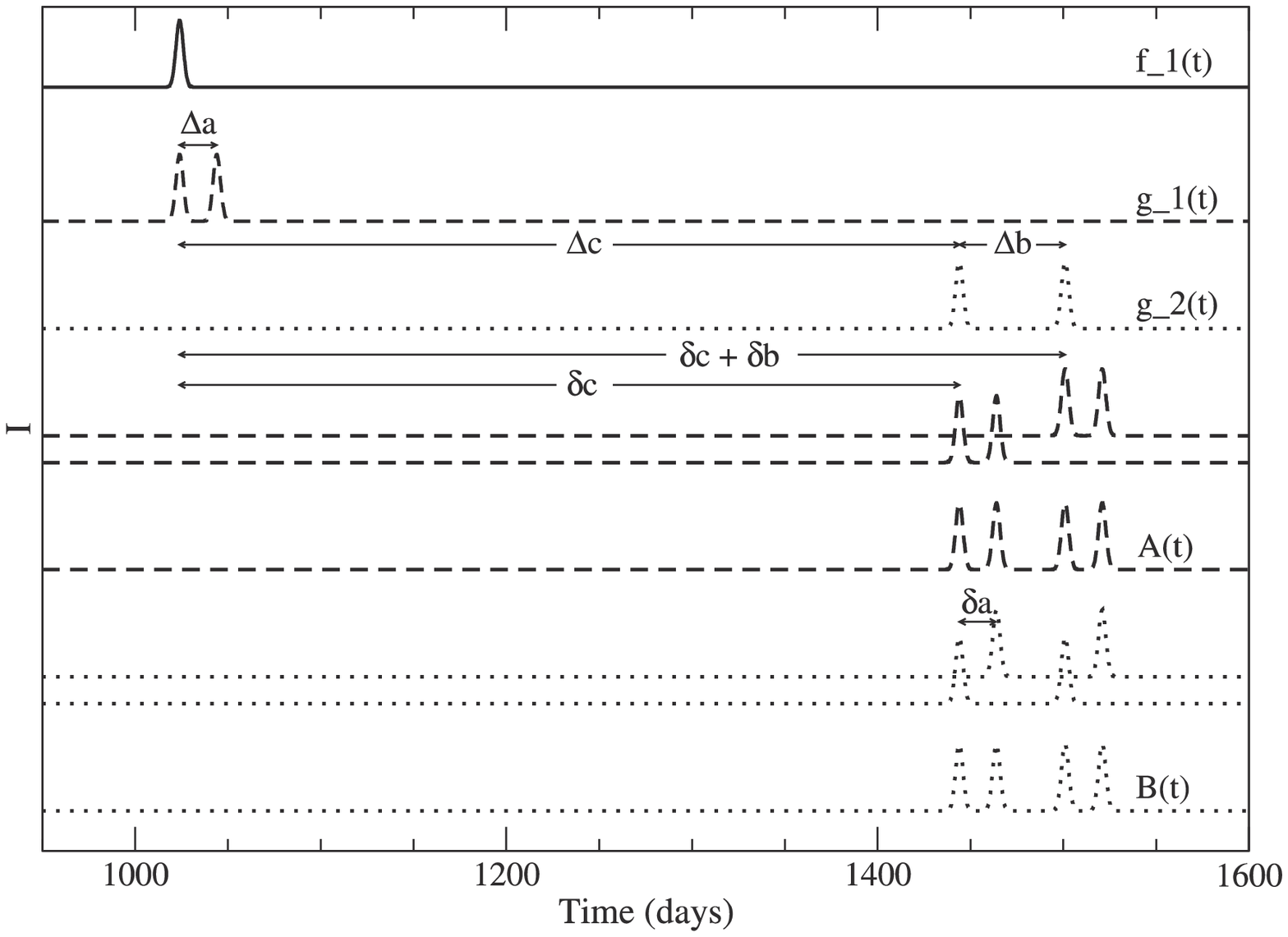,width=115truemm,angle=0,clip=}}
\captionc{1}{Graphical explanation of the method. See text for details.}
}
\end{figure}
\vspace{3mm}

\subsectionb{2.2}{The sampled case}

To build the combined sums from input data sequences (time, magnitude,
statistical weights) $t_{n}, g_{n}, W_{n}, n=1,2,...,N$ and the time
shifted versions of them $t_{m}, g_{m}, W_{m}, m=1,2,...,M$,
we form a table of triples
\begin{equation}
{t_n+t_m\over 2}, g_{n}+g_{m}, W_{n,m},
\label{k6verate_summa}
\end{equation}
using data from the sequence $g_{1}$ for blend $A$ and data from the
sequence
$g_{2}$ for blend $B$. Values $W_{n,m}$ are computed as
combined weights:
\begin{equation}
W_{n,m}=S_{n,m}{W_n W_m \over W_n+ W_m },
\label{summa_kaalud}
\end{equation}
where $S_{n,m}$ is the {\it downweighting function}:
\begin{equation}
S_{n,m}=\cases{1-{|t_n-t_m|\over \sigma },& if $|t_n-t_m|<\sigma $,\cr
               0,& if $|t_n-t_m|>\sigma$ \cr},
\label{sigmad}
\end{equation}
and $\sigma$ is the {\it downweighting parameter}, which depends on
sampling (it can be prefixed or chosen using trial calculations).

Next, by varying trial delays $\delta c$, $\delta a$ and $\delta b$
the artificial blends $A$ and $B$ are
recalculated and weighted sums of squared differences
between them are found:
\begin{equation}
D^2 =\min_{\alpha ,\beta} {\sum_{n,m} (\alpha A_n+\beta -B_m)^2
W_{n,m}^{A,B} \over \sum_{n,m} W_{n,m}^{A,B}}.
\label{distance}
\end{equation}

We may call $D^2$ as the {\it statistical distance} between $A$ and $B$.
Regression coefficients $\alpha$ and $\beta$ are needed, because
artificial blends may have different baselines and magnification.  They
are recalculated for every set of trial delays.  The combined weights
for $D^2$:
\begin{equation}
W_{n,m}^{A,B}=S_{n,m}{W_n^{A} W_m^{B} \over W_n^{A}+\alpha^2 W_m^{B}},
\label{vahe_ruutude_kaalud}
\end{equation}
are formed from $W_n^{A}$ and $W_m^{B}$ which are calculated using Eq.~\ref{summa_kaalud} for both artificial blends.

By varying trial delays $\delta c$, $\delta a$ and $\delta b$ over
pre-given grids, we are searching for the global minimum of statistical
distance $D^2$ which corresponds to the recovered time-delay system.
The weights contain parameter $\alpha$ to be estimated, consequently we
need an iterative scheme to get final dispersions.  First we set $\alpha
=1$ in Eq.~(\ref{vahe_ruutude_kaalud}), then solve linear weighted least
squares equations to get estimates for $\alpha$ and $\beta$.  After that
we insert a new value of $\alpha$ into weights and recompute.  Normally
this process converges in 3--4 steps.  A similar approach was used by
Hirv et al.  (2007).

\subsectionb{2.3}{Features and difficulties of the method}

There are three additional issues, which we have to bear in mind,
before starting actual calculations.

$\bullet$ Recovering the time-delay system is a degenerate problem.  The
mirrored values of short delays $\Delta a$ and $\Delta b$ are also
valid.  For a single data set we can get four equally correct solutions:
$\Delta c$, $\Delta a$ and $\Delta b$; $\Delta c+\Delta b$, $\Delta a$
and $-\Delta b$; $\Delta c-\Delta a$, $-\Delta a$ and $\Delta b$; and
$\Delta c-\Delta a+\Delta b$, $-\Delta a$ and $-\Delta b$.
(Interchanging $g_{1}$ and $g_{2}$ gives us four additional sets of
solutions, where $\Delta c$ is mirrored and $\Delta a$ and $\Delta b$
are interchanged.)  All the four solutions form detectable minima in the
three-dimensional grid of $D^2$ values.  For finite sequences these
minima may have slightly different merit function values.  Our method
just finds formally the deepest minimum and corresponding time-delay
system.  The recovered set of time delays may be considered real, if it
shows up as a visually noticeable minimum in the two-dimensional slice
of statistical distance values (see low-noise part of Figure~3).  Formal
significance estimation is possible using the bootstrap-type techniques
and ideas from Pelt et al.  (1996).

$\bullet$ Our method does not work if $|\Delta a|=|\Delta b|$.  Both
blends are then similar, and we can recover only the largest delay
$\Delta c$ using simplest ``one-dimensional'' dispersion spectra.
Having a value for the long delay it is then in principle possible to
recover the short delay (the same for both blends) from the combined
data using the methods described in Geiger \& Schneider (1996).  The
combining of two photometric series with estimated long delay allows
sometimes (if microlensing effect is negligible) to get a data set with
twice the original sampling rate.

The case of $|\Delta a|=|\Delta b|$ may be promptly recognized from the
plot of $D^2$ values -- one of the four possible solutions has a
characteristic distribution along straight line of $D^2$ values (see for
instance Figure~5).  We may also hit an arbitrary solution corresponding
to mirrored arbitrary short delays, which shows up as a normal minimum
in the two-dimensional plot of $D^2$ values.  Hence the solutions where
$|\delta a| \approx |\delta b|$ should be handled with care.  A
three-dimensional plot of $D^2$ values would be useful here.

The tests with simulated data-sets showed, that for a reasonably good
sampling and low noise even one day differences between $\Delta a$ and
$\Delta b$ values can be resolved.

$\bullet$ The process of calculating the $D^2$ values in the algorithm
for two blends is different from its analog for a clean curve and a
blend.  The calculation of $D^2$ involves differences of the observed
data sums.  In the case of a clean image and a blend, we have
differences of original data points and combined sums.  From what
follows that total scatter of the differences in the new method is
somewhat higher and statistical stability is lower.  Consequently, the
two blend method demands data with higher quality.

Our method for two blends recovers the time delays correctly also for a
clean image and a blend.  Because of different sensitivity to noise, it
is sensible to use proper method for the nature of a given problem.  For
unknown nature, it is worth trying both algorithms for the given input
data.

\subsectionb{2.4}{Data analysis}

The procedure of analyzing real data includes the following steps.  At
first, we need to find a suitable downweighting parameter $\sigma$ for a
given data sampling.  Next, we should verify if the noise level of our
data is under the noise value our method can handle.  And finally, we
may analyze the observed curves to recover the time-delay system.

We can use an interactive simulation for estimating the suitable
downweighting parameter for real observational data.  First, we generate
artificial noise-free curves with some pregiven time delays, using the
sampling of our real data.  Then starting from small downweighting
parameter (say $\sigma=0.5$), we move on towards larger ones and
recalculate the plot of $D^2$ and recover the time delays for each
$\sigma$.  In general, there is an optimal $\sigma$ for a given sampling
which recovers the time delays correctly and produces clearest minimum
on the $D^2$ surface.  Once we have found the optimum, further
enlargement of downweighting parameter will not improve the results.  It
is also possible, that for a given sampling and time-delay system, there
is no working downweighting parameter at all.  Even for a correctly
estimated value of $\sigma$ the overall success of the algorithm depends
on the length of the time series, noise level and absolute values of
actual time delays.  For the best results, $\sigma$ should not be larger
than half of the shortest time delay we are going to recover.  See also
Hirv et al.  (2007) where the estimation of the correct $\sigma$ is
presented in some detail.

To verify if the noise level of our data is tolerably low for the
method, we may use the same simulated curves that were used for
estimating the $\sigma$ parameter.  Taking these model sequences, adding
gradually increasing levels of Gaussian noise and recovering the
foreknown time delays, we can find the maximum tolerable noise level for
our algorithm.  Then we compare the signal to noise ratio ($S$/$N$) of
the observational data and test data which had the maximum tolerable
amount of noise.  The $S$/$N$ of observational data should be higher
than for the test data.

Once we have found that further analysis of the observational data is
reasonable, we perform the three-dimensional search for the minimum of
$D^2$.  The values of our trial parameters $\delta a$, $\delta b$
and $\delta c$ which correspond to the minimum of $D^2$, can be
considered as the real recovered time-delay system, if all the presented
above issues and complications were taken into account.

\sectionb{3}{GENERATING TEST DATA}

The quasar source variability $q_{n}, t_{n}$ is simulated using simple
random walk.  A randomly chosen value of $\pm 1.0$ is assigned
cumulatively to each step in the intensity scale.  The initial time
points are generated by using random step sizes from the interval
$[0.2,1.8]$ days.  Then the quasar signal is shifted in time by delays
$\Delta a$, $\Delta b$ and $\Delta c$ and blended as in the
Eqs.~(\ref{gg1}) and~(\ref{gg2}).  For blending the intensities, linear
interpolation is used.  The blend $g_{2}$ is multiplied by amplification
ratio $a=0.8$ to make things more realistic (the inherently important
assumption of the method is that both components of a given blend have
nearly equal magnification coefficients).  Both blends, $g_{1}$ and
$g_{2}$, can be resampled using generated or real observed sequences of
time points and linear interpolation.  Finally, the Gaussian noise is
added to the model observational noise.  One example of the generated
curves is shown in Figure~2.

\begin{figure}[!th]
\vbox{
\centerline{\psfig{figure=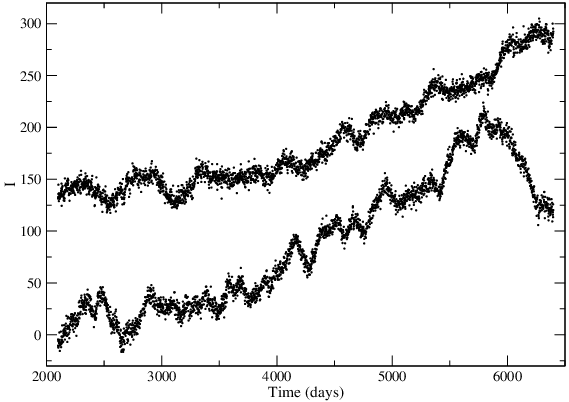,width=125truemm,angle=0,clip=}}
\vspace{-.5mm}
\captionb{2} {The computer-generated blends $g_{1}$ (lower curve) and
$g_{2}$ (shifted up by 130 units). $\Delta c=420.2$, $\Delta
a=20.2$, $\Delta b=56.5$ days  and $a=0.8$. The standard deviation of
the added Gaussian noise is 5 units.}
}
\vspace{5mm}
\end{figure}

\sectionb{4}{TESTING THE METHOD}

\subsectionb{4.1}{Simulated data}

Currently we do not have observational sequences at our disposal, that
are long enough, sampled well and have noise level our method can work
with.  So, to test the method, we had to build artificial sequences.  We
generated a $4300$ day long (2740 points) noise-free dataset with random
sampling which had only daylight caps; $\Delta c=420.2$, $\Delta
a=20.2$, $\Delta b=56.5$ days and $a=0.8$.  This set was also used for
estimating the optimal downweighting parameter for the given sampling.
Different levels of Gaussian noise were added to the computed curve to
check our method's stability against noise.  This set with added
Gaussian noise is shown in Figure~2.

\begin{wrapfigure}[16]{r}[0pt]{70mm}
\vspace{-.5mm}
 \parbox[c]{68mm}{ {\normbf \ \ Table 1.}{\norm\ Recovered time delays
depending on the added Gaussian noise.}}
\vskip1mm
\vbox{\small
\tabcolsep=15pt
\begin{tabular}{rrrr}
\tablerule
Noise & $\Delta c$ & $\Delta a$ & $\Delta b$ \\
\multicolumn{1}{r}{(stdev)}&
\multicolumn{3}{c}{} \\
\tablerule
$0.0$ & $400$ & $-20$ & $57$ \\[-2pt]
$1.0$ & $400$ & $-20$ & $57$ \\[-2pt]
$2.0$ & $401$ & $-19$ & $56$ \\[-2pt]
$3.0$ & $401$ & $-18$ & $56$ \\[-2pt]
$4.0$ & $401$ & $-18$ & $57$ \\[-2pt]
$5.0$ & $417$ & $19$ & $59$ \\[-2pt]
$6.0$ & $426$ & $27$ & $52$ \\[-2pt]
$7.0$ & $426$ & $19$ & $48$ \\[-2pt]
$8.0$ & $421$ & $26$ & $58$ \\[-2pt]
$9.0$ & $413$ & $20$ & $50$ \\[-2pt]
$11.0$ & $414$ & $42$ & $84$ \\[-2pt]
$14.0$ & $442$ & $43$ & $31$ \\[-2pt]
\tablerule
\end{tabular}
}
\end{wrapfigure}

For our test data the optimal downweighting parameter was
$\sigma=1.5$.  To find the maximally tolerable level of noise, we
performed a three-dimensional search for time delays, using the one day
step size and the following limits for trial delays:  $\delta
c=370...470$, $\delta a=-30...70$, $\delta b=6...106$.  The results of
noise tests are given in Table~1 and in Figure~3.  As we can see, the
Gaussian noise with a standard deviation of two units introduces one day
error in the estimates of $\Delta c$ and $\Delta a$; the Gaussian noise
with a standard deviation of six units gives us a seven day error in the
estimate of $\Delta a$; and a noise level of $11$ units makes an error
in the estimate of $\Delta a$ comparable to its original value.  The
signal to noise ratio was $S$/$N$ = 30 in the case of standard deviation
of six units.  Hence, for a reasonably well sampled real data we should
keep the $S/N \ge 30$ for the method to work properly.

\begin{figure}[!h]
\vbox{
\centerline{\psfig{figure=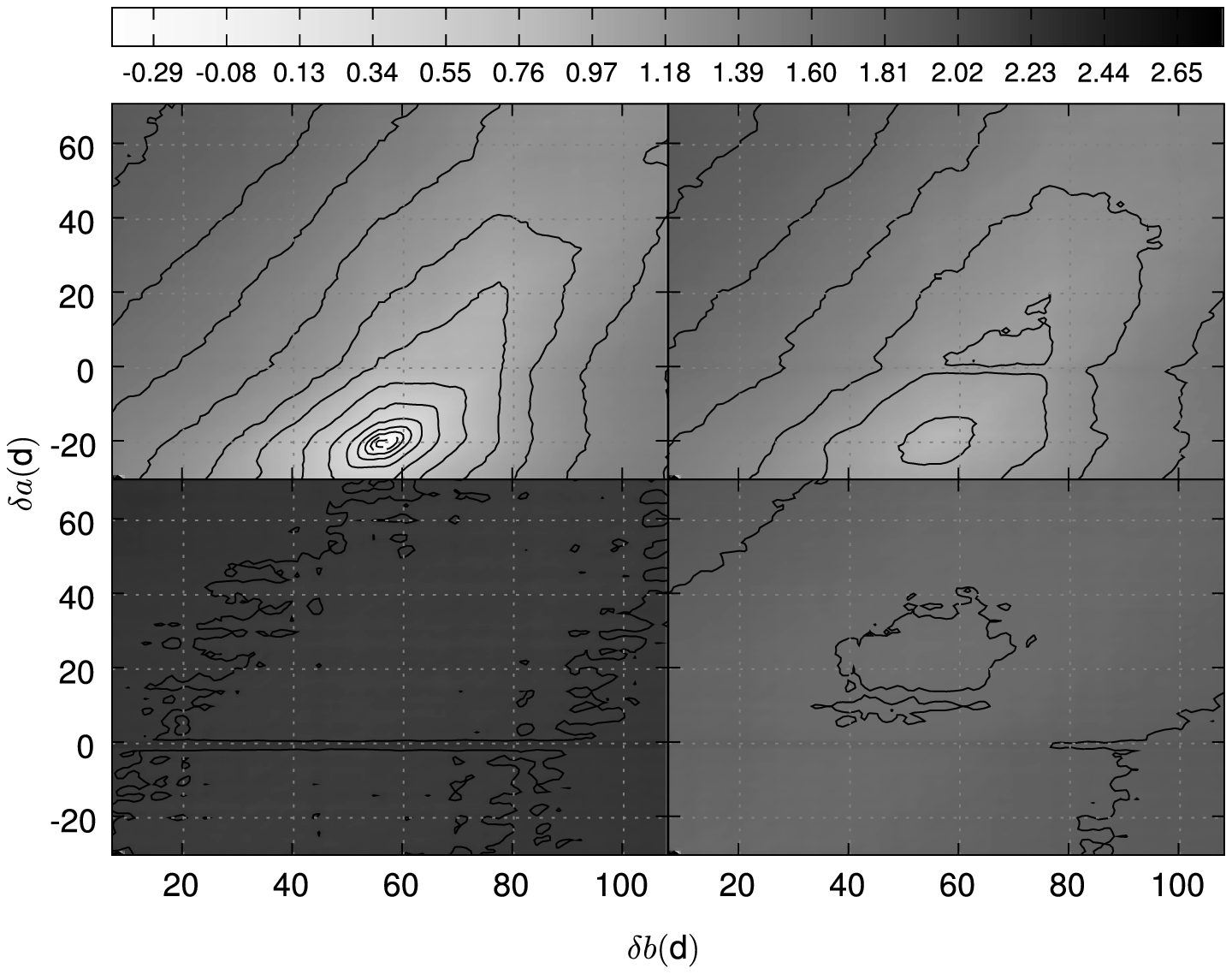,width=120truemm,angle=0,clip=}}
\vspace{-.5mm}
 \captionb{3}{Vanishing of the detectable minimum on the $D^2$ surface
due to observational errors. The standard deviations of  the
added Gaussian noise are (clockwise  from upper left): $0.0$, $2.0$,
$6.0$ and $11.0$. Values on the color key represent the
$log(D^2)$ and spacing of the contours.  The same type of color key is
used in all two-dimensional plots.}
}
\end{figure}

\begin{figure}[!h]
\centerline{\psfig{figure=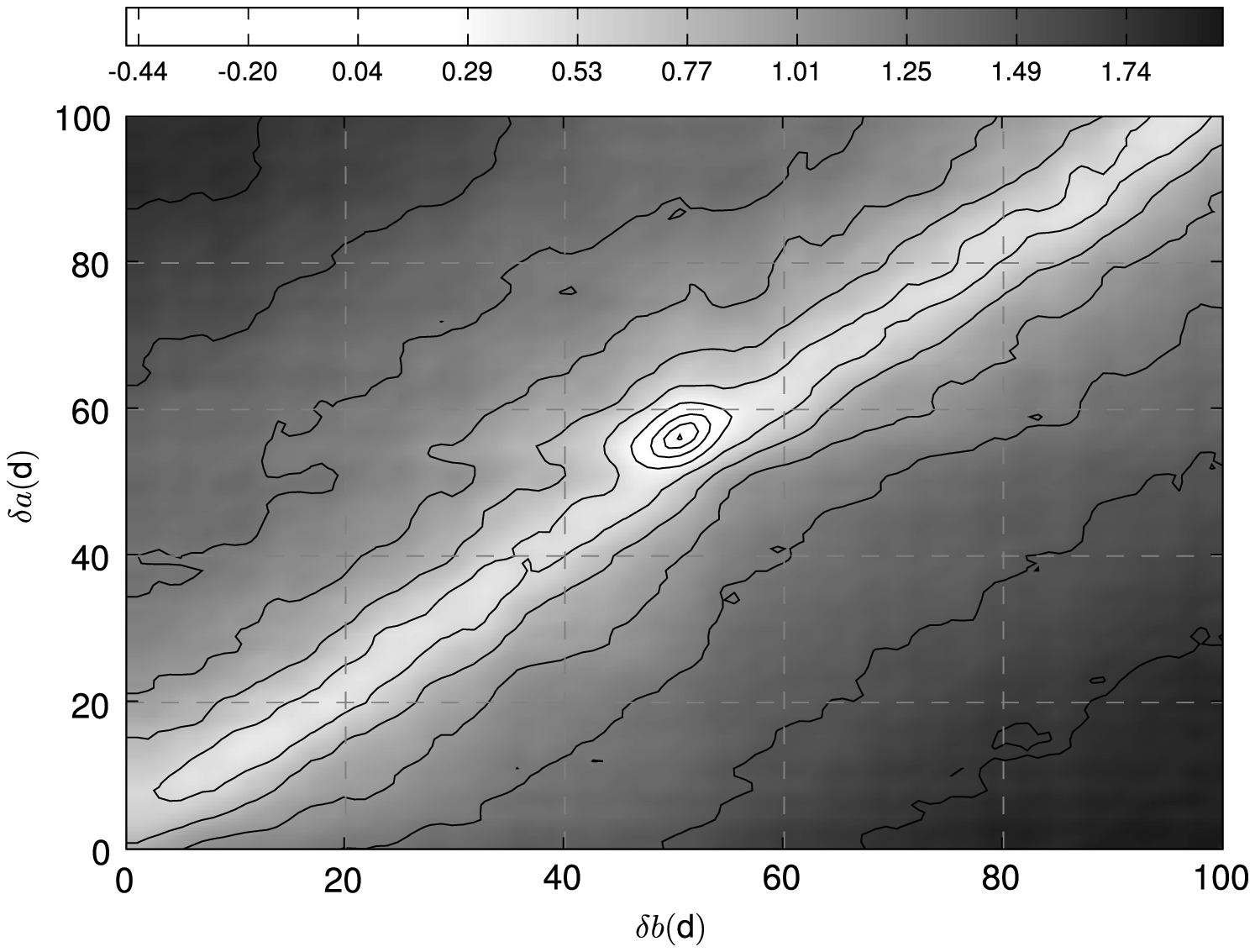,width=112truemm,angle=0,clip=}}
\vspace{-.5mm}
\captionb{4}{$D^2$ values  for very close short delays. $\Delta
c=420.2$, $\Delta a=56.5$, $\Delta b=50.1$ days and $a=0.8$.}
\vskip3mm
\centerline{\psfig{figure=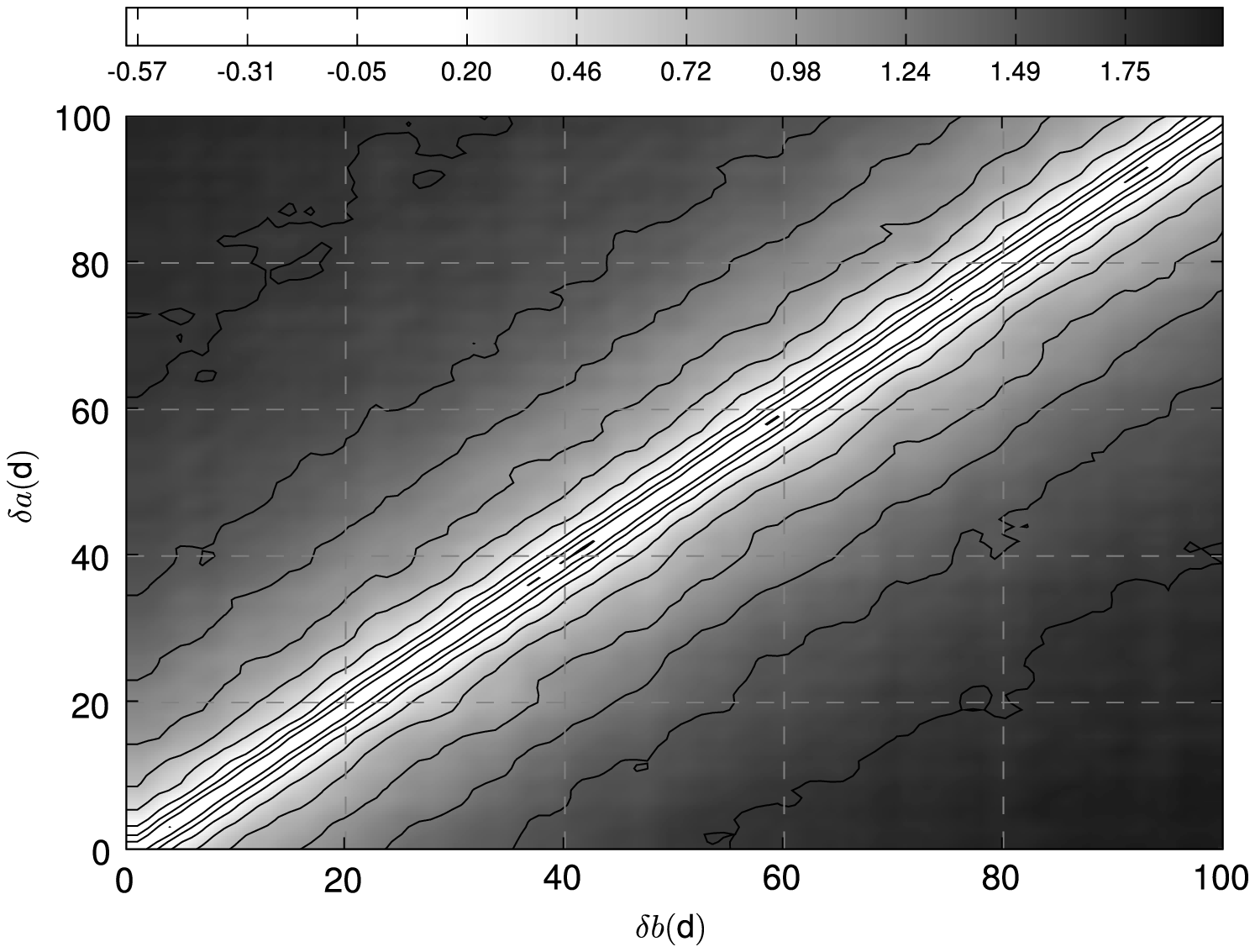,width=112truemm,angle=0,clip=}}
\vspace{-.5mm}
\captionc{5}{$D^2$ values  for $\Delta a=\Delta b=20.2$, $\Delta
c=420.2$ days and $a=0.8$.}
\end{figure}

\subsectionb{4.2}{Problem with Schild's data}

In Hirv et al.  (2007) we applied the algorithm for a clean image and a
blend to the observational data by Schild et al.  (1997).  As we got
then interesting results, we considered applying the algorithm for two
blends as well.  The optimal downweighting parameter $\sigma=1.5$ for
Schild's time series was found.  Next we found also the noise tolerance
of the method for two blends using Schild's sampling.  Having real
but bad sampling, where the points to days ratio is $0.2$ and large gaps
are included, the working noise tolerance decreased to $3.0$ units
(standard deviation).  The corresponding signal to noise ratio was
$S/N=50$.  Next we compared the signal to noise ratio of the Schild's
data and of our simulated curves.  Unfortunately the $S$/$N$ ratio for
Schild's data occurred to be lower than the ratio for good enough
simulated curves.  Because of that we cannot use assumption about two
blends for finding time delays from Schild's data.  Schild's noise level
was tolerable for the algorithm for a blend and a clean image, but not
for the algorithm for two blends.  (For the explanation see
Section~2.3.)

\subsectionb{4.3}{The $|\Delta a|\approx|\Delta b|$ case}

Having well sampled data and low noise, it is still possible to get a
solution for very close short delays.  For example, we took $\Delta
c=420.2$, $\Delta a=56.5$, $\Delta b=50.1$ days having a good sampling
with daylight caps only and no noise.  The given time delays were
recovered correctly.  The resulting plot of $D^2$ values is shown in
Figure~4.  Even a one day difference between $\Delta a$ and $\Delta b$
is still tolerable for the method, but then the minimum on the $D^2$
surface is not very convincing indeed.

The singular situation, where initial $\Delta c=420.2$, $\Delta a=20.2$,
$\Delta b=20.2$ days is shown in Figure~5.  We can see a characteristic
distribution along straight line of $D^2$ values and no minima, but this
is not always the case, as was discussed in Section~2.3.

\sectionb{5}{CONCLUSION}

A method for estimating time delays from two blended light curves was
developed and tested.  As the new algorithm is more sensitive to
observational errors than is the algorithm for a clean image and a
blend, we were not able to confirm the interesting results for real
Schild's data obtained in Hirv et al.  (2007).

Although we did not have observational data good enough for analysis,
all the steps of recovering time delays from real data were simulated as
realistically as possible.  For planning the real observations, one
should repeat similar simulations to establish realistic limits for
observational errors and sampling.

We believe that two new algorithms and the classical method of
dispersion spectra form a useful toolset to analyze data which will flow
out from the extensive photometric programs planned.  If enough data and
sufficient computing power will be available then we can set up a new
kind of searching program.  First, we select from a general database the
records, where two nearby measured images are variable, then we apply
the delay estimation schemes.  If we find that two curves can be
described as time shifted replicas of the single source curve or blends
with proper delay structure then the follow up spectroscopy can be
called upon.

\vskip5mm

ACKNOWLEDGMENTS.  This work was supported by the Estonian Science
Foundation grants Nos.~6810 and 6813.  Special thanks are to Krista
Alikas for valuable comments.

\vskip5mm

\References

\refb Burud~I., Magain~P., Sohy~S., Hjorth~J. 2001, A\&A, 380, 805

\refb Geiger~B., Schneider~P. 1996, MNRAS, 282, 530

\refb Gil-Merino~R., Wisotzki~L., Wambsganss J. 2002, A\&A, 381, 428

\refb Hirv~A., Eenm\"ae~T., Liimets~T., Liivam\"agi~L.~J., Pelt~J.
2007, A\&A, 464, 471

\refb Koptelova~E.~A., Oknyanskij~V.~L., Shimanovskaya~E.~V. 2006, A\&A,
452, 37

\refb Pelt~J., Hoff~W., Kayser~R., Refsdal~S., Schramm~T. 1994, A\&A,
286, 775

\refb Pelt~J., Kayser~R., Refsdal~S., Schramm~T. 1996, A\&A, 305, 97

\refb Pelt~J., Refsdal~S., Stabell~R. 2002, A\&A, 389, L57

\refb Cuevas-Tello~J.~C., Ti\v{n}o~P., Raychaudhury~S. 2006, A\&A, 454,
695

\refb Schild~R., Thomson~D.~J. 1997, AJ, 113, 130

\refb Vakulik~V., Schild~R., Dudinov~V. et al. 2006, A\&A, 447, 905

\end{document}